# Designing high-performance composite joints close to parent materials of aluminum matrix composites

Xuesong Leng[1], Wei Yang[1,2], Jiaheng Zhang[2], Xing Ma[1,2], Weiwei Zhao[1,2*], Jiuchun Yan[1*]

[1]*State Key Laboratory of Advanced Welding and joining, Harbin Institute of Technology, Harbin, 150001, People's Republic of China*

[2] *School of material science and technology, Harbin Institute of Technology (Shenzhen), Shenzhen, 518055, People's Republic of China*

∗*Ema*il: *wzhao@hit.edu.cn (WZ); jcyan@hit.edu.cn (JY)*

**Composite joints with high-performance close to the parent materials of aluminum matrix composites were fabricated by a new joining technology assisted by ultrasonic vibration. We found both the microstructure and the mechanical performance were systematically dependent on the volume fraction and the distribution of reinforcement particles in the bond region. This study can be generalized to the bonding of other ceramic-reinforced metal matrix composites.**

**Introduction**

Metal matrix composites (MMCs) were invented over decades[1, 2] and became more popular as structural materials[3-6]. Therefore the optimization of joining techniques is necessary for the application[7]. A number of joining techniques have been studied, including fusion welding (arc welding[8] and laser welding[9-11]), solid state bonding (diffusion bonding[12, 13] and friction stir welding[14, 15]), brazing[16], etc. However, there is still a common problem for these classical joining techniques: it is difficult to avoid the damage or the segregation of reinforcement particles in the bonding region during the bonding process. To bring ceramic particles into the bond region, a reactive diffusion joint was studied by directly mixing ceramic particles into aluminum alloy power as an interlayer [17]. However, the poor wettability of the cerimic particles with the alloy was unavoidable, and there were a number of porous zones in the bonds which were responsible for the low-performance of the joints[17]. Here, we report a new bonding technology to introduce ultrasonic wave into the diffusion bonding process to fabricate a composite structure bond with sufficient reinforcements, showing the high-performance close to that of the parent material.

In the fabrication methods for bulk MMC, many researchers have been encouraged to focus their attention to the use of ultrasonic wave[18, 19] since it can improve the wettability of the ceramic particles with the matrix alloy. Compared with the process to fabricate bulk MMC, besides of wetting problem between the ceramic particles with the matrix alloy, another challenge during the bonding process is how to avoid the segregation and get the uniform distribution of the ceramic particles in a very thin bond region. The recent research[20] exhibited ultrasonic treatment can improve the refinement of the matrix alloy grains and then make the ceramic particles distribution more uniform. Another advantage of this new technology is that there is no need to prefabricate the ceramic reinforcement particles in the initial filler material. By ultrasonic agitation and dissolution, sufficient reinforcements can be introduced into bond region from the MMC parent materials and a composite structure bond with sufficient reinforcements can be fabricated.

**Experiments**

The process of our joining technique is schematically shown in Fig.1. Liquid filler metal is able to fill into the gap between the two MMC parent materials by ultrasonic capillary effect[21] and an initial joint without ceramic particle forms as schematically shown in Fig. 1a. The liquid filler alloy can dissolve the MMC parent materials when the joint is heated to a certain higher temperature, and two dissolved layers form in the parent materials close proximity to the bond region as shown in Fig. 1b. After ultrasonic treatment is applied on the joint at this stage (Fig. 1c), the effect of the ultrasonic strongly improves the dissolution but also mix two dissolved layers with the bond region so that a new composite joint forms with uniformly distributed sufficient ceramic reinforcements (Fig. 1d).

MMC we used here is aluminum matrix composites $Al_{18}B_4O_{33}$/Al with shear strength of 240 MPa, which were fabricated by squeeze casting an Al-Cu-Mg alloy as matrix alloy with 20 volume percent $Al_{18}B_4O_{33}$ whiskers with an average diameter of 1 μm and a length of 20 μm[21]. Compared with the particle ceramic, the whiskers are beneficial for the studies of the anisotropic distribution under the ultrasonic treatment. The filler metal is a Zn-Al alloy with shear strength of 150 MPa.

**Results and discussion**

Fig. 2 shows the detailed experimental procedure on the joining technique. By 3s ultrasonic capillary effect at 400 ℃[21], the liquid Zn-Al filler alloy can fill into the gap between the two parent materials as shown in Fig. 2a. The amplitude and frequency of ultrasonic waves applied here are 20 μm and 20 kHz, separately. The thickness of the initial joint can be changed before bonding process by setting the gap between parent materials at two sides, but the joints with different thickness exhibit the same micromorphology[21]. In this work, we fixed the initial gay at 300 μm for all the studied joints. The microstructure of the bond region in Fig.2b shows a typical eutectic phase structure, which is consistent with that of the bulk Zn-Al filler alloy[20, 21]. The shear strength of the joint reaches to 145 MPa, which basically equals to that of the Zn-Al filler alloy, suggesting the performance of the initial joint is based on the ZnAl alloy itself since there is no any ceramic reinforcements in the initial bond.

In order to induce the ceramic reinforcements into bond region from MMC parent materials by the ultrasonic agitation, it is needed for the layers in parent materials close to the bond region to form the liquid state before the ultrasonic mixing process. According to the Zn-Al phase diagram[22], Zn-Al alloy will liquefy when Zn concentration beyond the liquid-solid phase boundary. At 500 ℃, the Zn-Al alloy will start to liquefy if the Zn concentration is increased over 42 wt %. As shown in Fig. 2c, the microstructure of the joint exhibits the thickness of the dissolved layer is about 150 μm after the initiate joint is heated to 500 ℃ and hold 5 min. There are three phases in the dissolved layer: whiskers (Black), Al-rich phase (Gray) and Zn-rich phase (white) as shown in Fig. 2d. The EDX measurement results show the Zn concentration in Al-rich phase and Zn-rich phase is 38.45 wt% and 42.69 wt%, respectively. Zn-rich phase are then the liquid phase at 500 ℃ and the dissolved layers are a semi-liquid-solid state. If the ultrasonic is applied at this stage as schematically shown in Figs. 1c-d, the liquid initial bond and two semi-liquid-solid dissolved layers in proximity will be mixed to form a new ceramic-reinforced bond by ultrasonic agitation. The microstructure of the joint (Fig. 2e) shows the thickness of the bond region become 600 μm after the ultrasonic process and whiskers are uniformly distributed in the whole bond. There is no whisker segregation (Fig. 2f) but also the whiskers are well bonded with the Zn-Al eutectic matrix (Fig. 2g).

By controlling the number of the ceramic particles which enter the bond region from the MMC parent material applying different technical parameters, we can control the fraction of the ceramic whiskers in the bond region as shown in Fig. 3. As anticipated, the mechanical performance of the joints increases with the increasing of the whisker fraction in the bond region. This reinforcement-fraction-dependent behavior is same as that of the MMC bulk[23], suggesting the new technology studied here for the MMC bonding exploits the same reinforced mechanism for the MMC parent materials. When the whisker fraction in the bond reaches 20 percent by volume, the shear strength is increased to ~220 MPa, 92 % of the parent material value. Another interesting behavior is also observed: the grain size of the matrix alloy are refined with the increasing of the whisker, that is also helpful to improve the performance of the joint. This can be explained that the growth of the crystal grains of the eutectic matrix is interrupted during cooling process by the presence of large numbers of whiskers in the bond region[24]. Fig. 2g shows the whiskers extend into the Zn-Al eutectic phase with laminae of 100 nm (eutectoid phase grows alternatively in the white α-Al phase and the black β-Zn phase) while the parallel growth orientation of the laminae is blocked by the presence of the whisker.

**Conclusions**

In summary, by a new developed joining technology based on ultrasonic bonding and diffusion bonding, we can design the composite joints with the various fraction of the reinforcement particles in the bond region. When the fraction of the whiskers reach 20 % (same as the parent materials), the performance is close to that of the parent material. This approach can be used for bonding of other kinds of ceramic-reinforced metal matrix composites, for example, SiC (or $Al_2O_3$) ceramic reinforced Al (or Cu) matrix composites.

**Acknowledgments**

This research was sponsored by the Natural Science Foundation of China (No.50375039) and supported by the Excellence Team Program at the Harbin Institute of Technology.

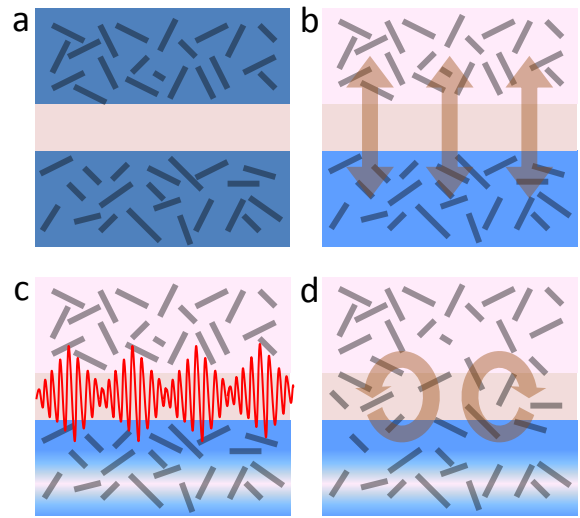

**Fig.1 The schematic of the joining technology process**. (a) Liquid filler metal is able to fill into the gap between the two MMC parent materials and an initial joint without ceramic particle forms by ultrasonic capillary effect[21]. (b) The liquid filler alloy can dissolve the MMC parent materials and dissolved layers forms in the MMC close proximity to the bond region after the joint is heated to a certain higher temperature. (c) Ultrasonic treatment is applied on the joint at the stage of (b). (d) A new composite joint forms with uniformly distributed sufficient ceramic reinforcements.

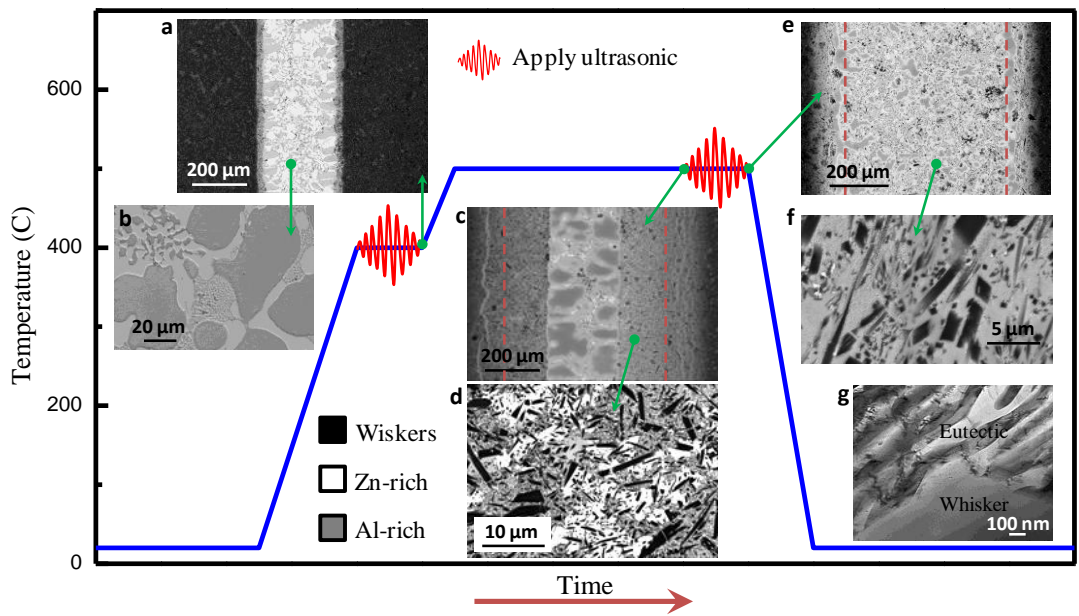

**Fig. 2. The bonding procedure and microstructures of the joints at key stages**. (a) A SEM image of the initial joint of 150 μm thickness without any ceramic reinforcements, which forms by ultrasonic capillary effect[21]. (b) A zoom-in SEM image of the bond region showing the microstructure is consistent with bulk Zn-Al filler metal. (c) A SEM image showing two dissolved layers with the thickness of 150 μm in proximity to the parent materials form after the initiate joint is heated to 500 ℃ and hold 5 min. (d) A zoom-in SEM image showing three phases in the dissolved layer: whiskers (Black), Al-rich phase (Gray) and Zn-rich phase (white). Zn-rich phase are liquid at 500 ℃. (e) A SEM image of the bond region after applying 3s ultrasonic on the bond shown in (c). The bond region becomes 600 μm after the ultrasonic process and whiskers are uniformly distributed in the whole bond. (f) A zoom-in SEM image showing no whisker segregation in the composite bond. (g) A TEM image showing whiskers are well bonded with the Zn-Al eutectic matrix.

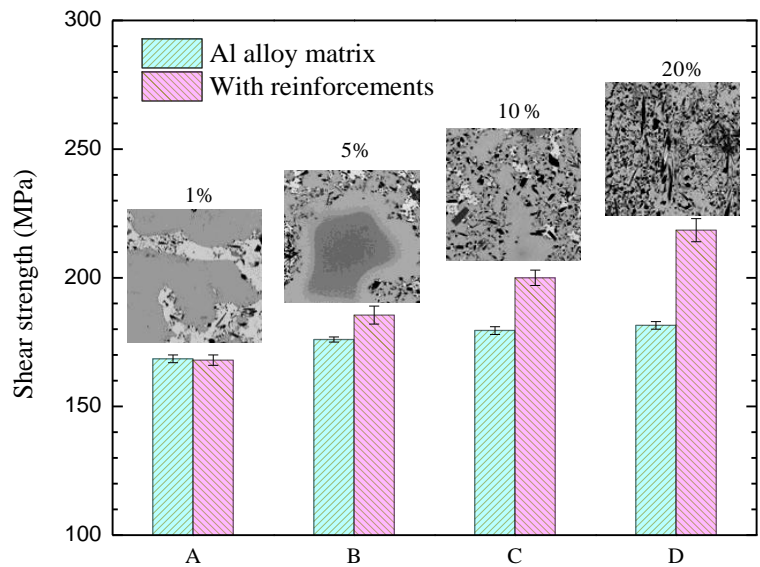

**Fig. 3. A serial of designed composite bonds by applying different technical parameters.** The whisker fraction in the bond region is 1%, 5%, 10%, 20% by volume in technologies of Mode A,B,C,D, respectively. The shear strength of the joints increases with the increasing of the whisker fraction in the bond region. The highest strength of the joint with 20 % whisker fraction is ~220 MPa, that reaches to 92 % of the parent material value. In four modes, all parameters are same except of the condition when ultrasonic agitation is applied. Mode A: ultrasonic agitation is applied when the initiate joint is heated to 450 ℃ without pressure on the two parent materials. Mode B: 450 ℃, with 2 MPa pressure. Mode C: 500 ℃, without pressure. Mode D: 500 ℃, with 2 MPa pressure. The pressure can help to squeeze the liquid matrix alloy out of the bond region and keep whiskers in the bond so that the whisker fraction in the bond can be increased.